\documentclass[useAMS,usenatbib]{mn2e}
\newcommand{\oiii}{\>{\rm [O}\,{\sc III}{\rm ]}}   
\newcommand{\oii}{\>{\rm [O}\,{\sc II}{\rm ]}}     
\newcommand{\nii}{\>{\rm [N}\,{\sc II}{\rm ]}}     

\newcommand{\beq}{\begin{equation}}
\newcommand{\eeq}{\end{equation}}

\newcommand{\apj}{ApJ}
\newcommand{\apjl}{ApJL}
\newcommand{\apjs}{ApJS}

\newcommand{\mnras}{MNRAS}
\newcommand{\aap}{A\&A}

\newdimen\hssize
\newdimen\hdsize
\usepackage{amssymb} 

\usepackage{graphicx}

\title[The MZ relation for local SFGs] {The MZ relation for local star-forming galaxies}
\author[Wu et  al.]{Yu-Zhong Wu$^{1,2}$\thanks{E-mail: yzwu@nao.cas.cn},
       Shuang-Nan Zhang$^{1,3}$,
       Yong-Heng Zhao$^{1,2}$,
       Wei Zhang$^{1,2}$\\
       ${^1}$  National Astronomical Observatories, Chinese Academy of Sciences, Beijing 100012, China\\
       ${^2}$  Key Laboratory of Optical Astronomy, National Astronomical Observatories, Chinese Academy of Sciences, Beijing 100012, China\\
       ${^3}$  Key Laboratory for Particle Astrophysics, Institute of High Energy Physics, Chinese Academy of Sciences, 19B Yuquan Road, \\~~~~Beijing 100049, China}

\begin{document}

\date{Accepted ........ Received ........; in original form ........}

\pagerange{\pageref{firstpage}--\pageref{lastpage}} \pubyear{2010}

\maketitle

\label{firstpage}

\begin{abstract}

We investigate the evolution of the mass-metallicity (MZ) relation
with a large sample of 53,444 star-forming galaxies (SFGs) at
$0.04<z<0.12$, selected from the catalog of MPA-JHU emission-line
measurements for the Sloan Digital Sky Survey (SDSS) Data Release 7.
Regarding the sample of SFGs, we correct the observational bias and
raise the aperture covering fractions to check the reliability of
the metallicity evolution. We show that (1) the redshift evolution
of log($\rm H\alpha$) and log($\rm \oiii$) luminosities is displayed
in our sample; (2) we find the metallicity evolution of $\sim 0.15$
dex at $\rm log (M_{*}/M_{\sun})\sim9.3$ in SFGs at $0.04<z<0.12$;
(3) after applying the luminosity thresholds of log$(L_{\rm H
\alpha})>41.0$ and log$(L_{\rm \oiii})>39.7$, we find that
metallicity evolution is shown well, and that SFR evolution still is
shown well under the latter luminosity threshold, but the evolution
is not observed under the former one; (4) the evolution of the MZ
relation seems to disappear at about $\rm log(M_{*}/M_{\sun})>10.0$
after applying the luminosity threshold of log$(L_{\rm H
\alpha})>41.0$ or log$(L_{\rm \oiii})>39.7$; (5) we find $\alpha
=0.09$ and $\alpha =0.07$ in the equation ($\mu={\rm
log}M_{*}-\alpha \rm log(SFR)$) for log$(L_{\rm H \alpha})>41.0$ and
log$(L_{\rm \oiii})>39.7$ samples, respectively, and these imply
that the evolution of the MZ relation may have a weaker dependence
on SFR in our sample.

\end{abstract}


\section{INTRODUCTION}

The gas-phase oxygen abundance (metallicity) is a key parameter to
understand the processes of the formation and evolution of a galaxy,
because metallicity mirrors both the history of gas inflow and
outflow and the result of the past star-forming activity. In most
star-forming galaxies (SFGs) at z $\lesssim 2$, observations suggest
that the stellar mass ($M_{*}$) increase originates mainly from
cosmological inflows of gas from the intergalactic medium (Noeske et
al. 2007; Whitaker et al. 2012; Zahid et al. 2013). Meanwhile,
outflows of gas which exist almost ubiquitously can be observed out
to z $\thicksim$ 3 in SFGs (Weiner et al. 2009; Steidel et al.
2010). Because metallicity is established by the interaction of gas
flows and star formation, this implies that a relation may exist
between the galaxy stellar mass and metallicity.

A measurement of the average metallicity using as a function
 of stellar mass is called as the mass-metallicity (MZ) relation.
The presence of an MZ relation for SFGs was first observed by
Lequeux et al. (1979), and following observations found that the
relation of SFGs is generally shifted to lower metallicities with
redshift (e.g. Erb et al. 2006; Maiolino et al. 2008; Zahid et al.
2013), compared with the $z = 0.1$ relation of Tremonti et al.
(2004, hereafter T04). Recently, Juneau et al. (2014) reported that
the evolution of the observed MZ relation may be dependent on sample
selection and the threshold line luminosities, and that the
emission-line luminosity limits can bias our perception of the
metallicity evolution of SFGs; applying a higher luminosity
threshold, the local MZ relation becomes steeper. Using the star
formation rate (SFR) as the third parameter, Mannucci et al. (2010)
proposed the fundamental metallicity relation (FMR), the relation
between Mass-Metallicity-SFR, which has been studied and discussed
(Lara-L\'{o}pez et al. 2010, 2013; Mannucci et al. 2010; Yates,
Kauffmann, \& Guo 2012; Henry et al. 2013b).

With the developments of near-infrared spectroscopy surveys, the MZ
relation has been measured at many galaxy samples with different
redshifts (Savaglio et al. 2005; Erb et al. 2006; Maiolino et al.
2008; Mannucci et al. 2009; Henry et al. 2013a, 2013b; Cullen et al.
2014; Steidel et al. 2014; Salim et al. 2015). Henry et al. (2013a)
used 26 galaxies at $z \sim 0.6-0.7$ to present the relation with
$0.12$ dex decrease in metallicity compared with the local relation,
and showed that they follow the local FMR. Utilizing 93 galaxies at
$z\gtrsim2$, Cullen et al. (2014) presented that the $\sim 0.3$ dex
lower metallicity in these galaxies than the local FMR may be an
artefact of the MZ evolution, based on different metallicity
indicators and calibrations. For these high redshift samples which
are restricted to high emission-line luminosities, investigating the
line luminosity effect in the Sloan Digital Sky Survey (SDSS) sample
will help us understand how to interpret high-redshift SFG
observations. Although the relation for SFGs at various redshift
ranges has been extensively discussed, and much progress has been
made, we will use the SFG sample at lower redshifts ($z < 0.12$),
which is considered the effect of line luminosity evolution, to
investigate the relation.

In this paper, we investigate the MZ relation for SFGs at $z<0.12$,
selected from the catalog of MPA-JHU measurements for the SDSS Data
Release 7 (DR7). In Section 2, we describe mainly the SFG sample
obtained by a series of steps. In Section 3, we present and discuss
the evolution of the MZ relation in the local SFGs. We summarize the
results and conclusions in Section 4. We adopt $H_0=70~\rm
km~s^{-1}~Mpc^{-1}$, $\Omega_{M}=0.3$, and $\Omega_{\Lambda}=0.7$.

\section{THE SAMPLE and DATA}

To explore the MZ relation at $z<0.12$, we take into account the
effect of line luminosity evolution and the aperture covering
fraction, and compile a sample, selected from the catalog of the
SDSS MPA-JHU DR7 release; 927,552 spectra are included in the
measurements.

A number of parameters, such as various emission line measurements,
are provided in the release. Based on the BPT diagram (Baldwin et
al. 1981; Kauffmann et al. 2003; Kewley et al. 2006), we use the
condition

$$\rm log([O~III]/H\beta)<0.61/(log(N~II)/H\alpha -0.05)+1.3)$$ to
obtain the SFG sample of 369,479 galaxies. Considering the bias of
the MZ relation from the aperture effect, a lower redshift limit of
0.04 is required (Kewley et al. 2005), and an upper redshift limit
of 0.12 is chosen to minimize evolution effect (Zahid et al. 2013,
2014), and therefore it can avoid or reduce a selection effect of
metallicity evolution originating from higher redshift galaxies. Due
to the above redshift cuts, our sample leaves 199,645 galaxies.
Considering the bias of the aperture effect, we can remove galaxies
whose aperture covering fractions are $<20\%$ for r band; 110,261
galaxies are removed from the sample.

Because high metallicity galaxies often have very weak $\rm
\oiii~\lambda 5007$ emission (Zahid et al. 2014), SNR cuts do not
employ the $\rm \oiii$~$\lambda 5007$ lines to avoid the bias of the
measured MZ relation (Foster et al. 2012); therefore we choose
galaxies from the above sample with S/N (SNR) $>$ 3 for
H$\beta~\lambda 4861$, H$\alpha~\lambda 6563$, $\rm \oii$~$\lambda
\lambda 3727,3729$, and $\rm \nii$~ $\lambda 6584$. Using the SNR
cut, our sample has 57,321 galaxies. Finally, we have chosen these
galaxies with both $M_{*}
>0.0$ and an SFR FLAG keyword of $0$. Our final sample contains
53,444 SFGs.

In addition to the measurements of the redshifts, we also use the
stellar masses and SFRs provided by the MPA-JHU group. The stellar
masses are obtained, following the philosophy of Kauffmann et al.
(2003) and Salim et al. (2007); the SFRs are based on Brinchmann et
al. (2004); the stellar masses and SFRs assumed a Kroupa (2001)
initial mass function (IMF) are corrected using a Chabrier (2003)
IMF. Due to no SFR error provided by the MPA-JHU catalogue, we use
$\rm 0.5\times(SFR_{84}-SFR_{16})$ as the error (Lian et al. 2015),
where $\rm SFR_{84}$ and $\rm SFR_{16}$ are the 84th and 16th
percentile values of log(SFR) in the catalogue.

\begin{figure}
\begin{center}
\includegraphics[width=8cm,height=6cm]{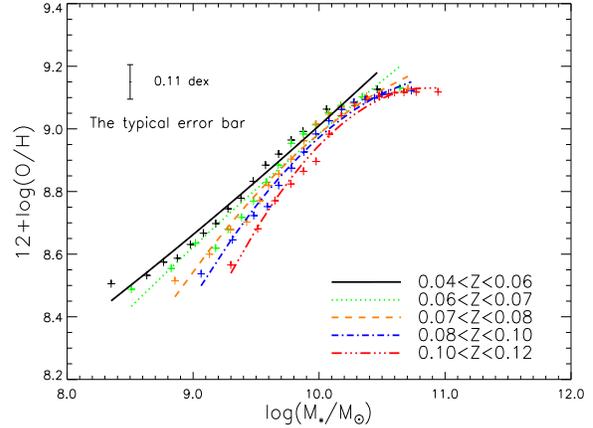}
\caption{Mass-metallicity relations for SFGs in different redshift
subsamples. ``+'' signs show the median metallicities and stellar
masses sorted into 16 bins for each subsample, with more than $100$
galaxies in each bin. The different lines are the second-order
polynomial fits for SFGs with different redshift ranges. $0.11$ dex
in Figure 1 is the typical error bar.}
\end{center}
\end{figure}


Oxygen abundances of SFGs are estimated using the $R_{23}$ method
(Pilyugin et al. 2006, 2010; Wu \& Zhang 2013) and we adopt the
calibration of T04 in this paper. To break the degeneracy between
the upper and lower branch $R_{23}$ solutions, we choose the SFGs
with log($\rm \nii \lambda6584/\rm \oii\lambda3727$) $>-1.2$ as our
sample (Kewley \& Ellison 2008), and we find that all SFGs of our
sample have $\rm log([NII]/[OII])>-1.2$.

\section{Results}

In this section, we first investigate whether there is any redshift
evolution of the MZ relation within our SFG sample. We then explore
whether there is a subsequent evolution in emission line
luminosities for log(H$\alpha$) and log(H$\alpha$), and how this
effects the MZ evolution. We then investigate the SFR dependence of
the MZ relation in our sample. Finally, we investigate the biases
related to variations in the aperture covering fraction within the
sample.

\begin{figure*}
\begin{center}
\includegraphics[width=7cm,height=5.25cm]{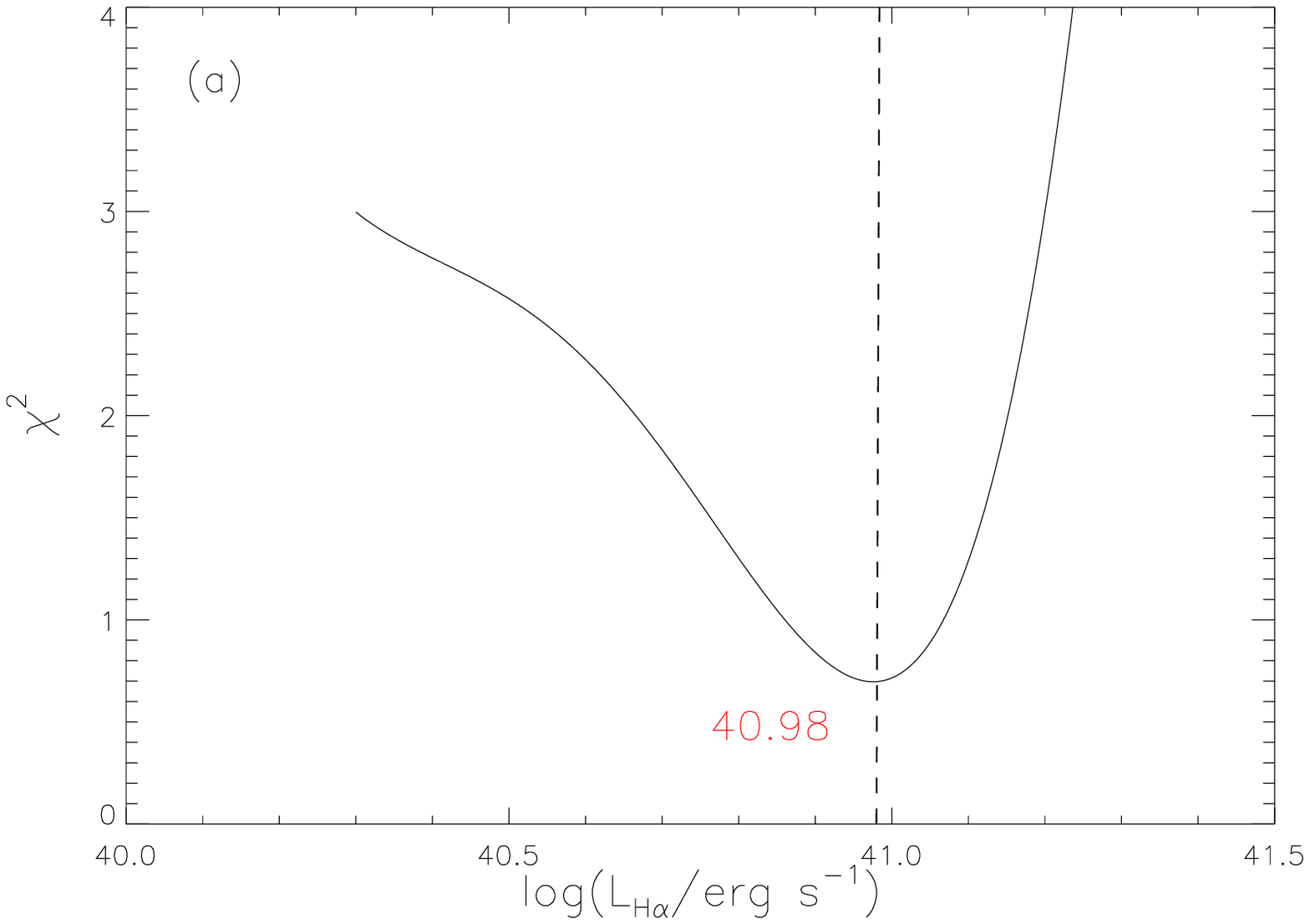}
\includegraphics[width=7cm,height=5.25cm]{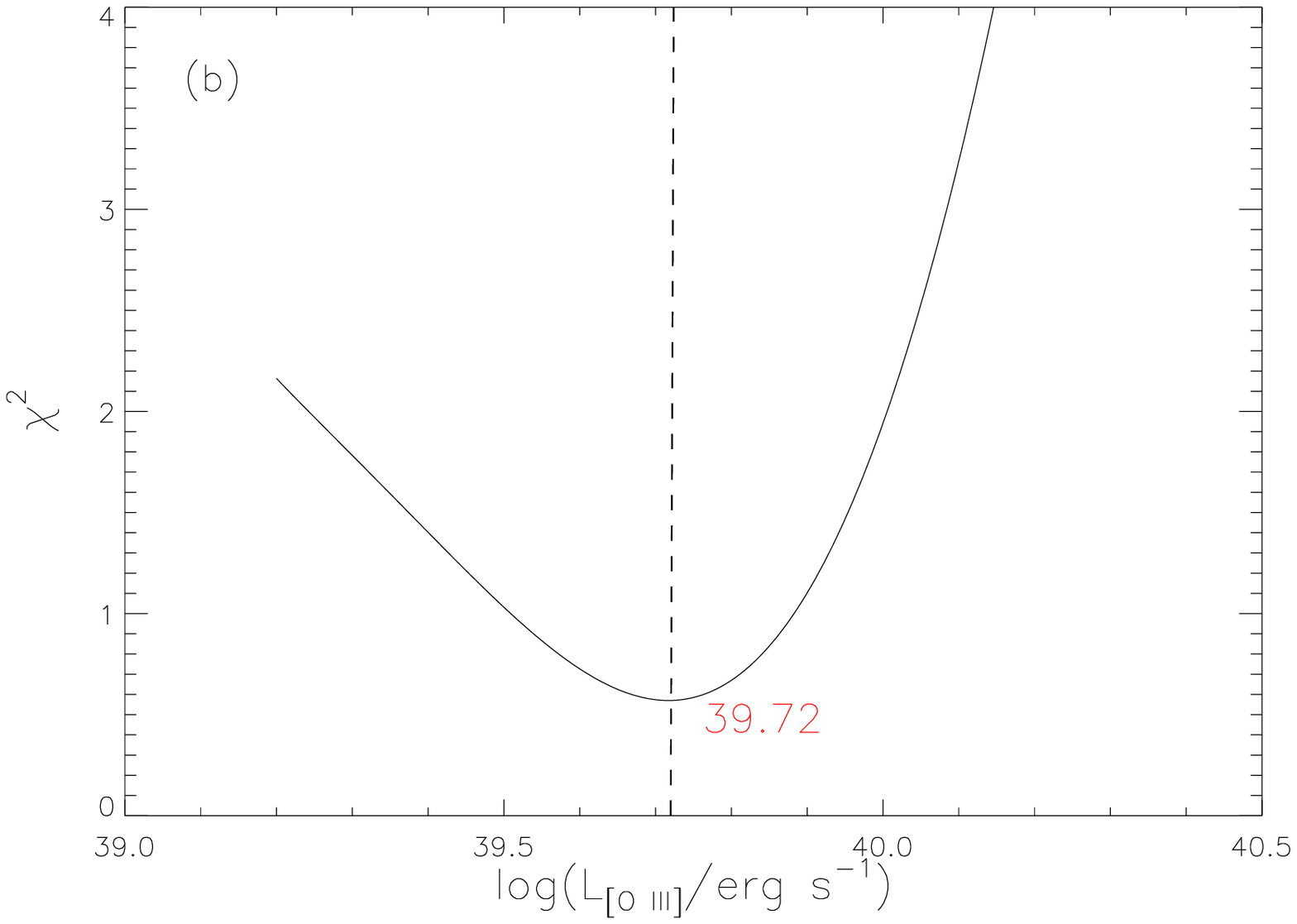}
\includegraphics[width=7cm,height=5.25cm]{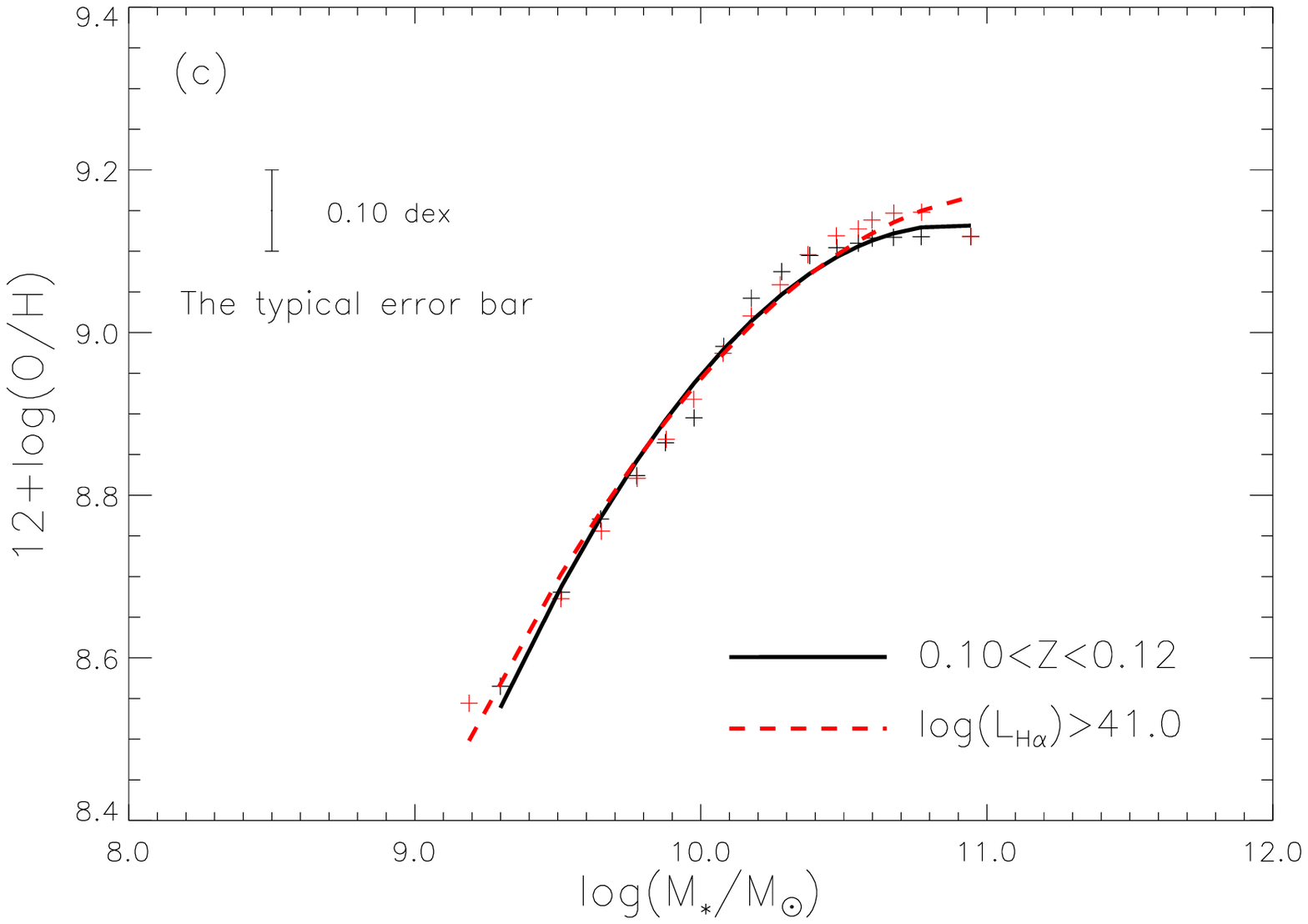}
\includegraphics[width=7cm,height=5.25cm]{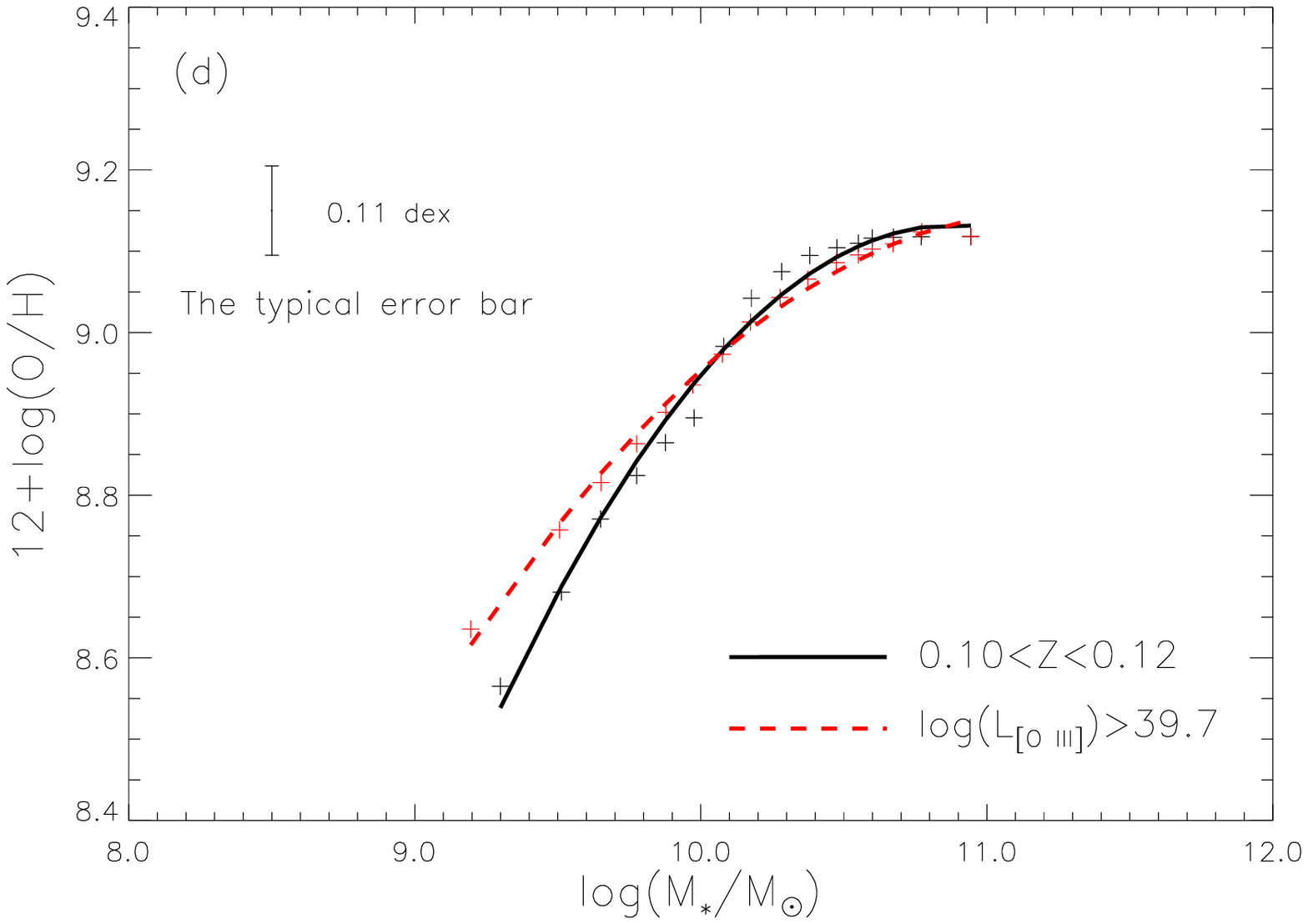}
\caption{The minimum luminosity limits of log($L_{\rm H\alpha}$) and
log($L_{\rm \oiii}$). The luminosity thresholds of log($L_{\rm
H\alpha}$) and log($L_{\rm \oiii}$) are derived using a Chi-squared
test to assess the minimization of $\chi^2$ in the MZ relations
between SFGs at $0.10<z<0.12$ and above the different luminosity
limits.}
\end{center}
\end{figure*}

\begin{figure*}
\begin{center}
\includegraphics[width=7cm,height=5.25cm]{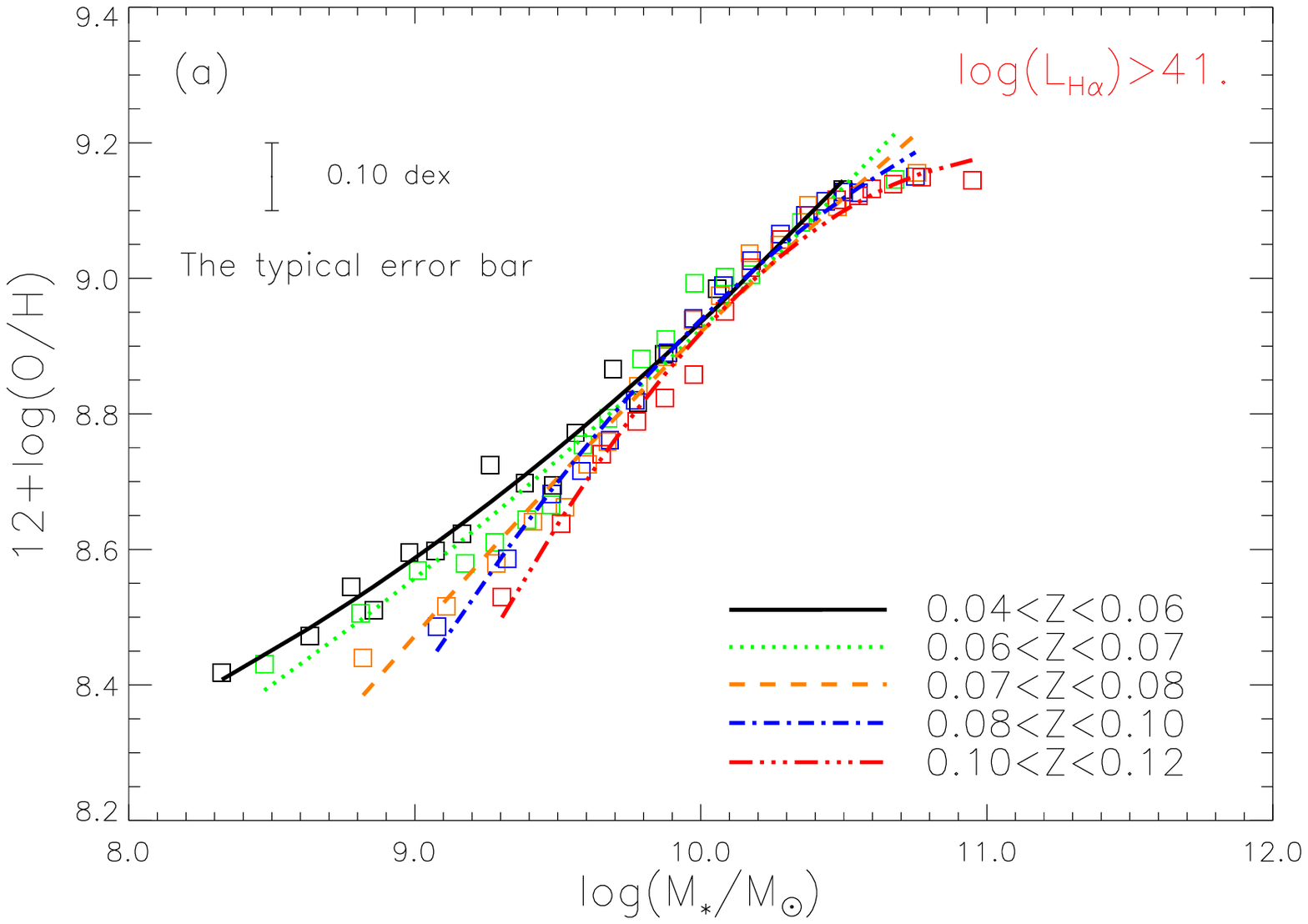}
\includegraphics[width=7cm,height=5.25cm]{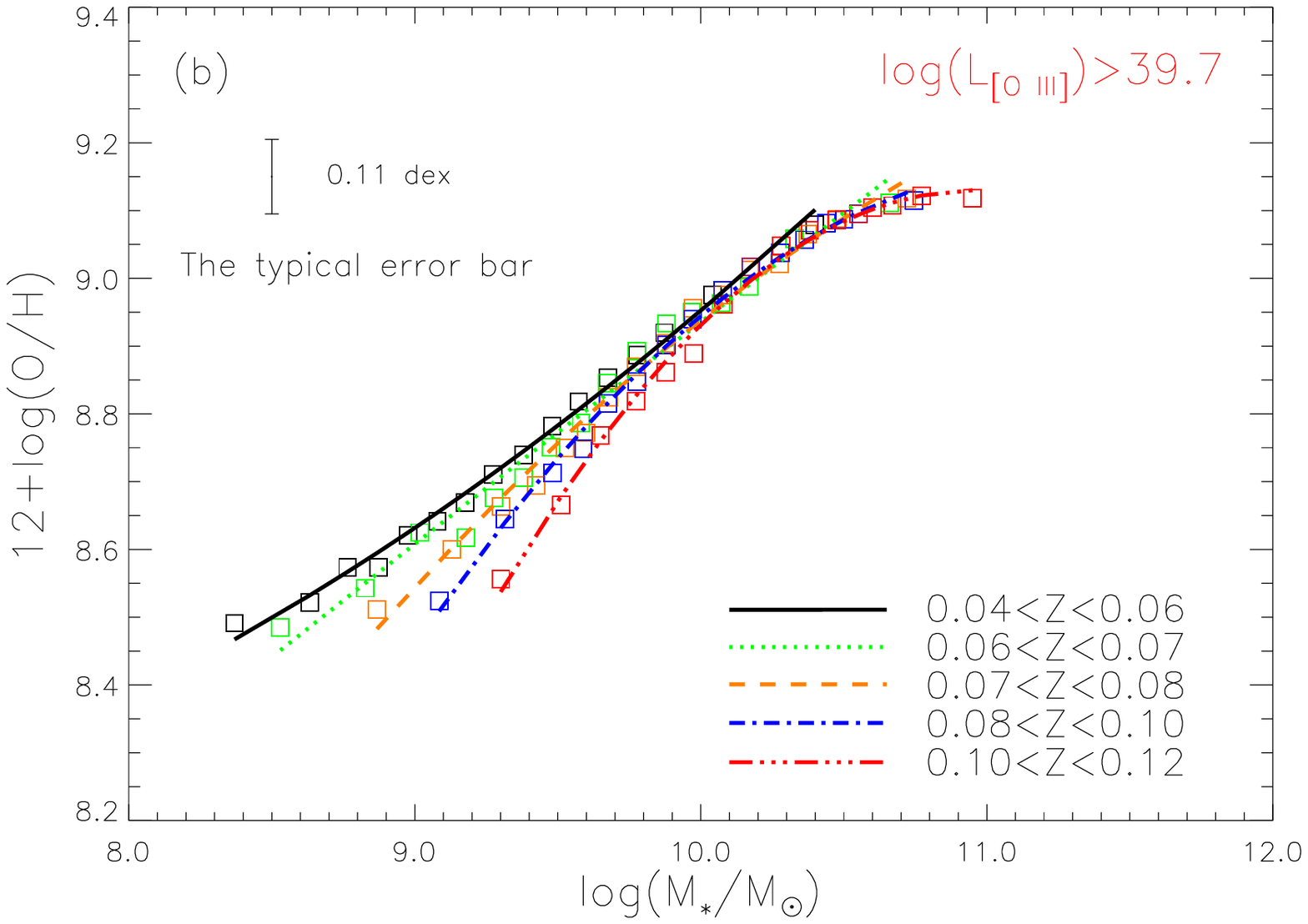}
\includegraphics[width=7cm,height=5.25cm]{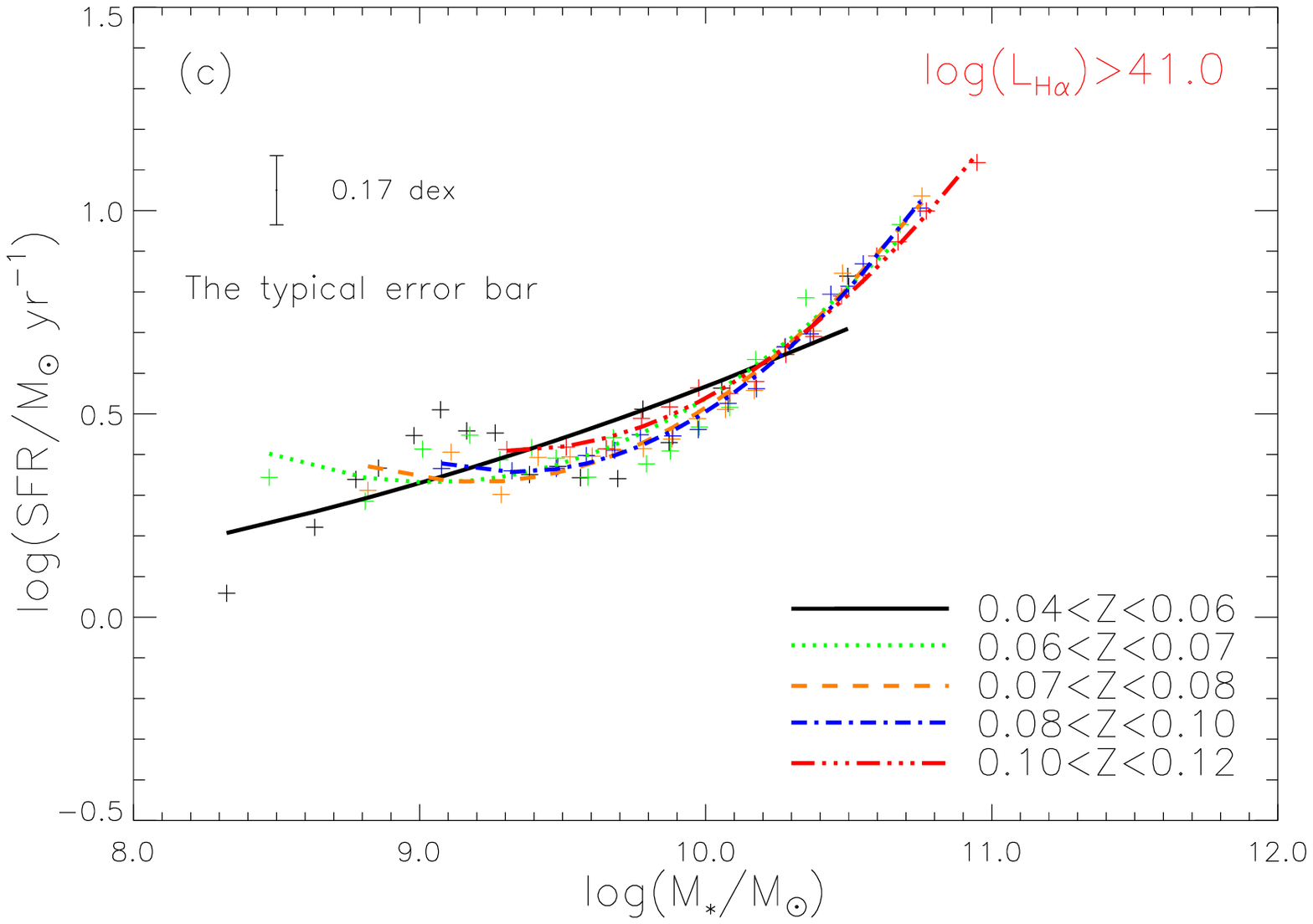}
\includegraphics[width=7cm,height=5.25cm]{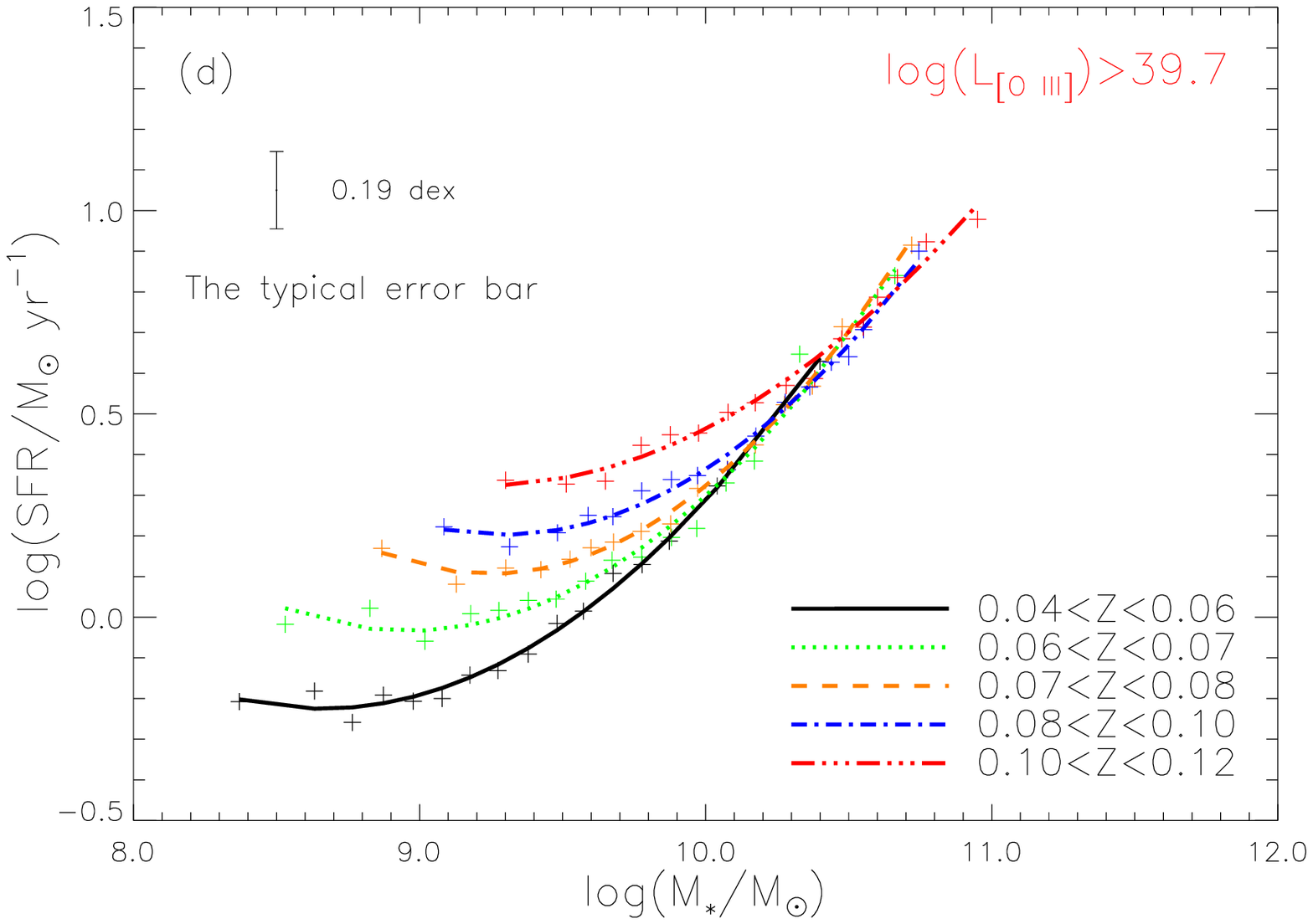}
\caption{Metallicity (open square dots) and SFR (`+' dots)
evolutions for the different minimum luminosity limits of
log($L_{\rm H\alpha}$) and log($L_{\rm \oiii}$). The MZ relations
with the typical error bars of $0.10$ dex and $0.11$ dex in Figures
3(a) and 3(b) are displayed. The M-SFR relations with the typical
error bars of $0.17$ dex and $0.19$ dex in Figures 3(c) and 3(d) are
displayed. The different lines show a second-order polynomial fit to
their local SDSS data. The stellar mass bins are the same as in
Figure 1.}
\end{center}
\end{figure*}

\begin{table*}
\caption{The redshift subsample parameters of the local SFGs.}
\begin{small}
\begin{center}
\setlength{\tabcolsep}{7.5pt}
\renewcommand{\arraystretch}{1.0}
\begin{tabular}{ccccccl}
\hline \hline  Redshift ranges& log(SFR)[$\rm M_{\sun}~yr^{-1}$] &
log($L_{\rm H \alpha}$)[$\rm erg~s^{-1}$] & log($L_{\rm \oiii})[\rm
erg~s^{-1}$] & z
 & N &   Total
 \\
 &Median &Median & Median & Median
 &   &
 \\
(1)& (2) & (3) & (4) &(5)&(6) &(7)\\
\hline
 $0.04<z\leqslant0.06$        &    -0.05 &    40.31  & 39.42  & 0.050  & 13370   &         \\
 $0.06<z\leqslant0.07$        &     0.14 &    40.52  & 39.51  & 0.065  & 8017    &         \\
 $0.07<z\leqslant0.08$        &     0.25 &    40.62  & 39.55  & 0.075  & 8760    & 53444   \\
 $0.08<z\leqslant0.10$        &     0.39 &    40.75  & 39.64  & 0.089  & 12929   &         \\
 $0.10<z<0.12$                &     0.56 &    40.90  & 39.76  & 0.110  & 10368   &         \\
\hline \hline
\end{tabular}
\parbox{6.5in}
{\baselineskip 11pt \noindent \vglue 0.5cm {\sc Note}: Col.(1):
redshift ranges. Col.(2)-(5): the median values of log(SFR),
log($L_{\rm H \alpha}$), log($L_{\rm \oiii})$, and redshifts,
respectively. Col.(6): the size of each subsample. Cols.(7): the
total sample size.}
\end{center}
\end{small}
\end{table*}

\subsection{Emission line luminosity evolution and metallicity
evolution}

Based on our SFG sample, we sort it into 5 subsamples; their
redshift ranges are $0.04<z\leqslant0.06$, $0.06<z\leqslant0.07$,
$0.07<z\leqslant0.08$, $0.08<z\leqslant0.10$, and $0.10<z<0.12$, and
we show their median values of $\rm log(SFR)[M_{\sun}/yr]$ and
redshifts in Table 1. Table 1 shows clearly increasing SFR with
redshift.

To explore well the MZ relation for SFGs with different redshift
ranges, we sort SFGs of each subsample into 16 bins of galaxy
stellar mass; more than 100 galaxies are included in each bin. All
median values of stellar masses and metallicities of SFGs in each
bin are shown in Figure 1, and the typical error bar of $0.11$dex in
metallicity is presented in Figure 1. In addition, we employ the
second-order polynomial to fit the data of each subsample, and
present them with different lines. In Figure 1, the metallicity
difference of the MZ relations between the lowest and highest
redshift ranges at $\rm log (M_{*}/M_{\sun})\sim9.3$ is about $0.2$
dex (the red line and black line in Figure 1). In addition, Figure 1
shows that the metallicity difference is more significant toward
lower stellar masses, suggesting that this may be a consequence of
the downsizing effect of metal-enriched gas, i.e., higher mass
galaxies evolving at earlier times on the MZ relation than lower
mass ones; this agrees with the result of Henry et al. (2013b) and
Yabe et al. (2014).

Using the SDSS sample, an evolution of log($\rm H\alpha$) and
log($\rm \oiii$) luminosities was found by Juneau et al. (2014). In
our redshift subsamples, we investigate whether there is a redshift
evolution of log($\rm H\alpha$) and log($\rm \oiii$) luminosities.
In Table 1, we list their median values of log$(L_{\rm H\alpha})$,
log$(L_{\rm \oiii})$, and redshifts. The evolution of log($\rm
H\alpha$) and log($\rm \oiii$) luminosities is clearly shown.

\begin{figure*}
\begin{center}
\includegraphics[width=7cm,height=5.25cm]{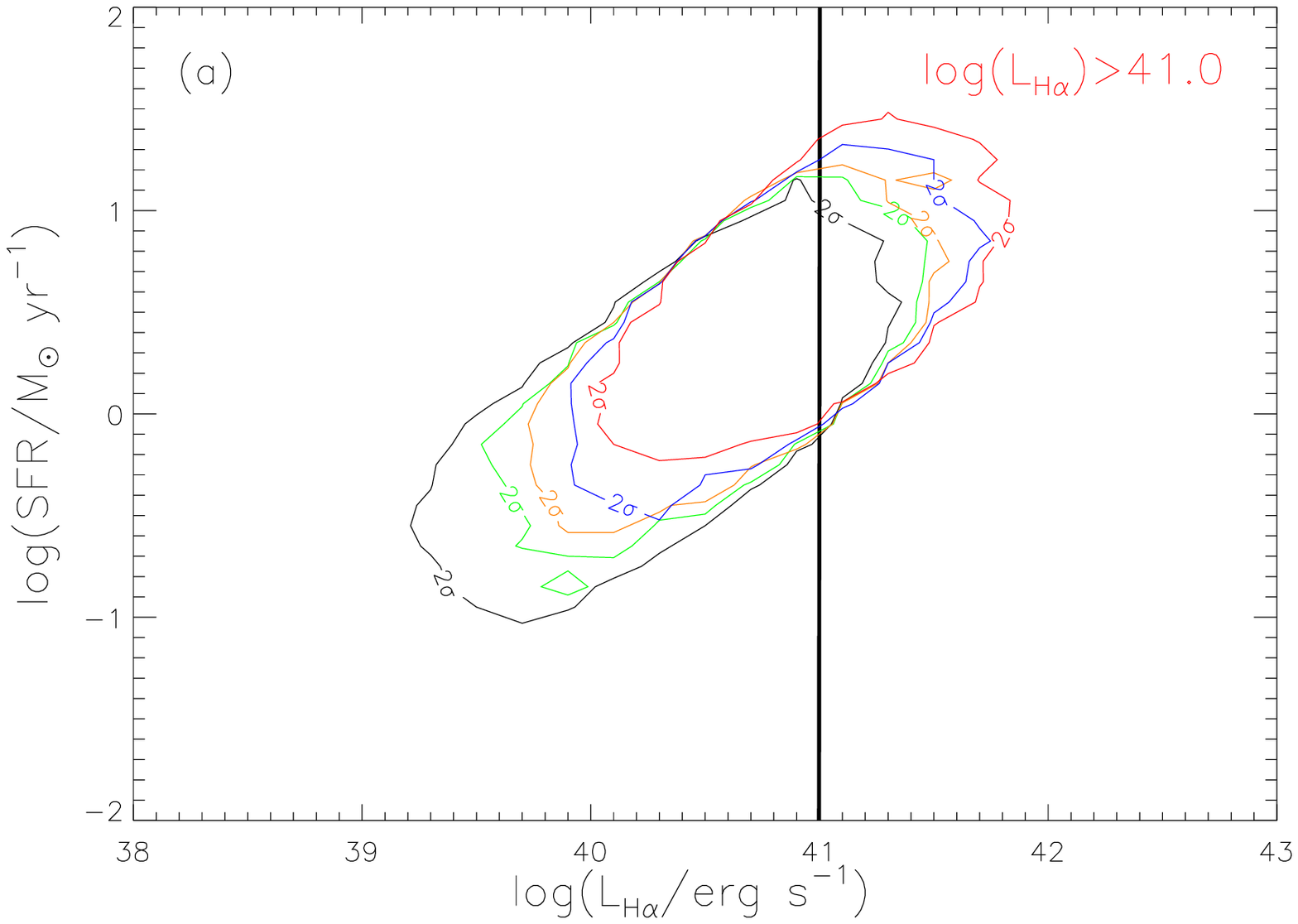}
\includegraphics[width=7cm,height=5.25cm]{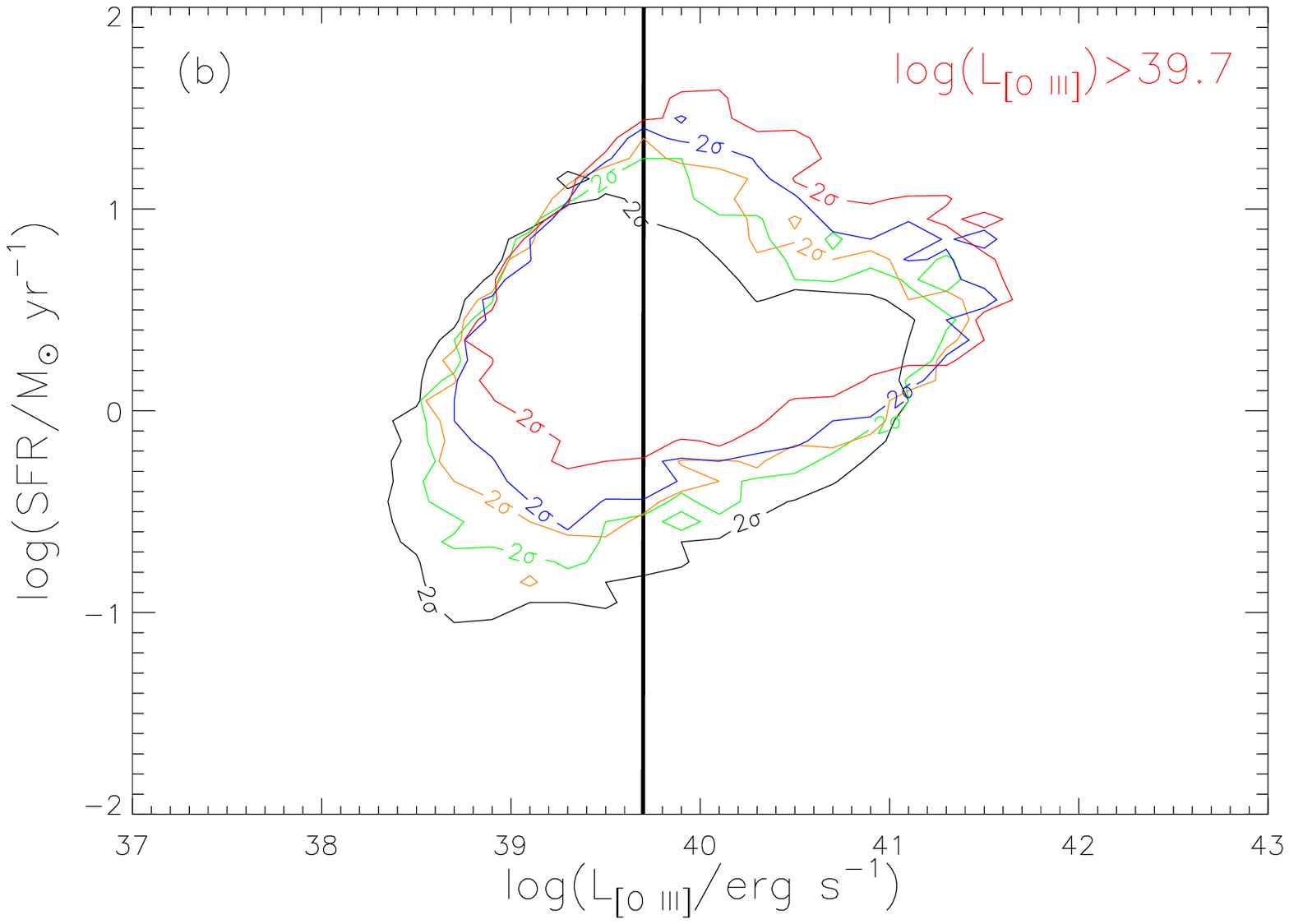}
\includegraphics[width=7cm,height=5.25cm]{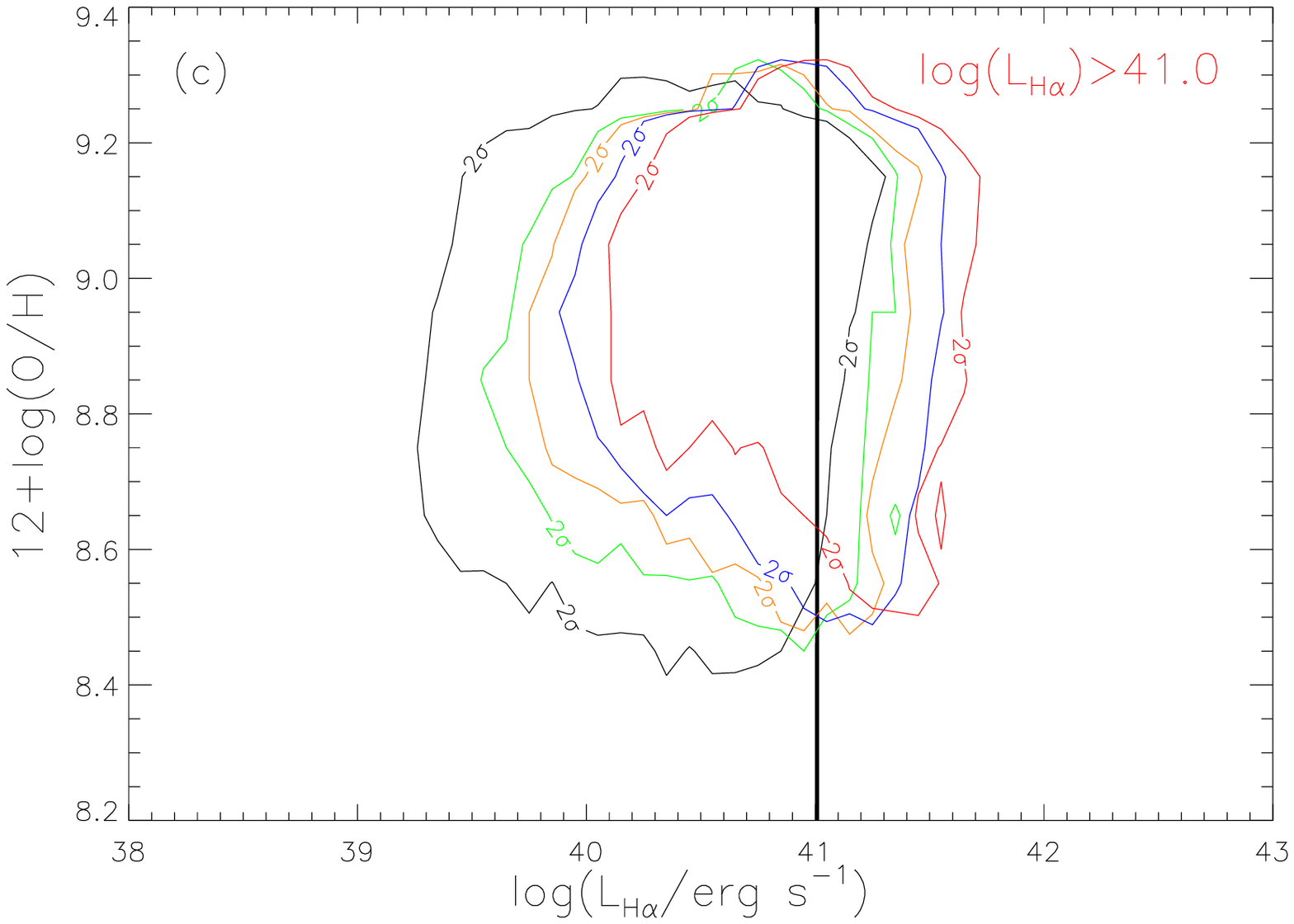}
\includegraphics[width=7cm,height=5.25cm]{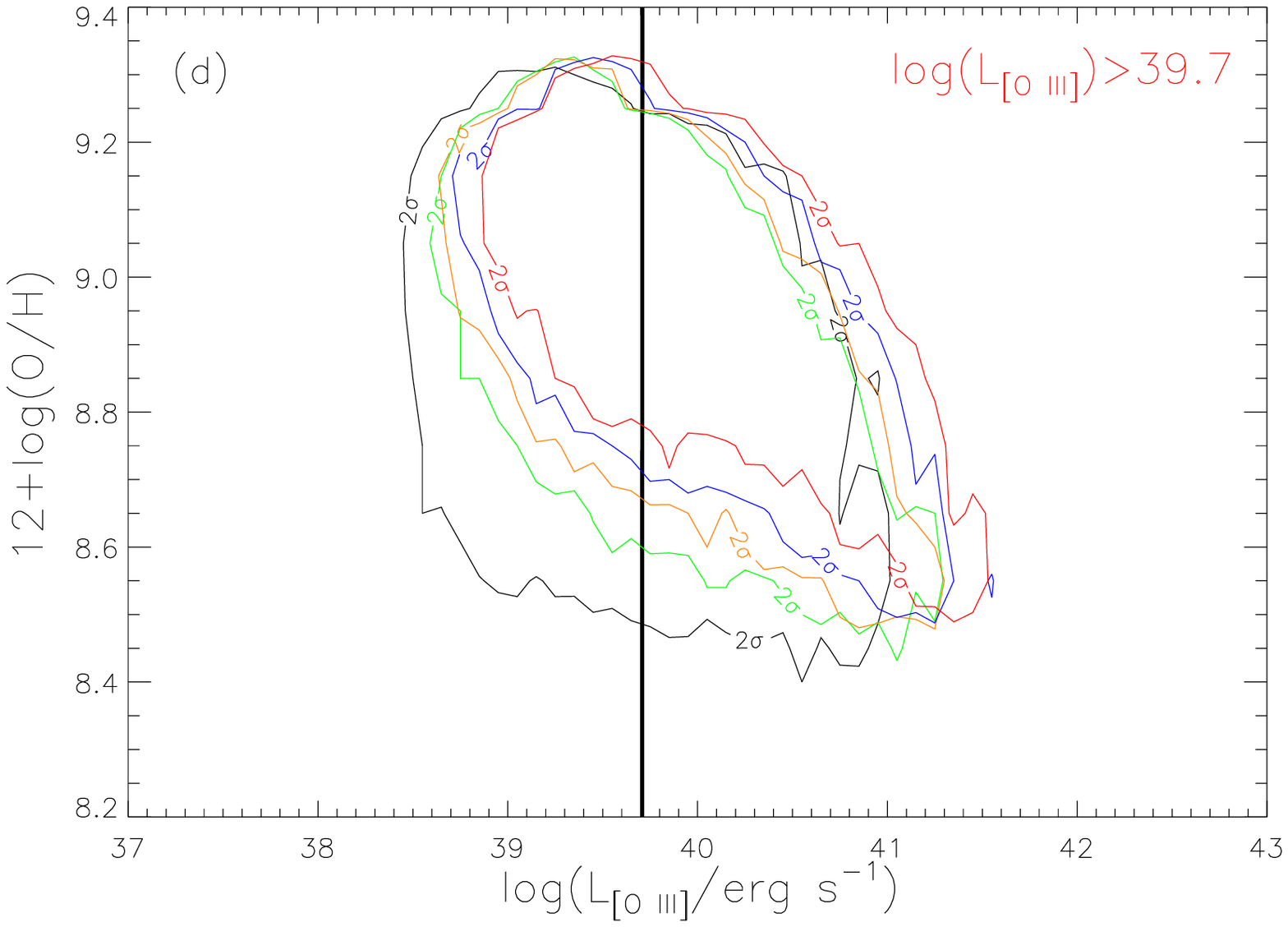}
\caption{Contours of SFRs or 12+log(O/H) used as a function of
log($\rm H\alpha$) and log($\rm \oiii$) luminosities for these
different redshift SFGs (the different colour contours correspond to
their different redshift samples of Figure 1). $2 \sigma$ shows the
$2\sigma$ regions of the Gaussian distributions for both SFRs or
12+log(O/H) and log($\rm H\alpha$) or log($\rm \oiii$) luminosities.
The straight line in each panel is the threshold luminosity of log($
L_{\rm H\alpha}$) or log($L_{\rm \oiii}$).}
\end{center}
\end{figure*}

Utilizing the log($\rm H\alpha$) and log($\rm \oiii$) luminosity
limits for the SDSS sample, Juneau et al. (2014) showed an
artificial evolution of MZ relation. The metallicity evolution shown
in Figure 1 also should be corrected by the observed minimum line
luminosity limit. We employ the MZ relation of the $0.10<z<0.12$
subsample to correct and achieve the observed minimum line
luminosity limit in our $0.04<z<0.12$ sample; then the lower
redshift subsamples can be cut to include only galaxies above that
line luminosity limit. A Chi-squared test is used to assess the
minimization of $\chi^2$ in the MZ relations between SFGs at
$0.10<z<0.12$ (the black solid lines of Figures 2(c) and 2(d)) and
above the different log($\rm H\alpha$) or log($\rm \oiii$)
luminosity thresholds. Considering the metallicity difference
between the above mentioned two samples, we first obtain the median
metallicity of each galaxy stellar mass bin in the log$(L_{\rm
H\alpha})>41.0$ or log$(L_{\rm \oiii})>39.7$ SFG sample and the
metallicities at the median stellar masses of the above mentioned
stellar mass bins, which are given by a linear interpolation of the
metallicities in the MZ relations of the $0.10<z<0.12$ subsample;
then we summate their metallicity differences between the two
samples at the same stellar masses in 14 galaxy stellar mass
bins\footnote{In the lowest and largest galaxy stellar mass bins of
the $0.10<z<0.12$ subsample (the first and sixteenth mass bins),
since the metallicities at their corresponding median galaxy stellar
masses can not be obtained by the interpolation method, we obtain
the total metallicity difference from the other 14 galaxy stellar
mass bins (from the second to the fifteenth mass bins), when
calculating the chi-squared values.}; finally, we obtain
respectively the luminosity thresholds of log$(L_{\rm H
\alpha})>40.98 \approx 41.0$ and log$(L_{\rm \oiii})>39.72 \approx
39.7$ (Figures 2(a) and 2(b)).

\begin{figure*}
\begin{center}
\includegraphics[width=7cm,height=5.25cm]{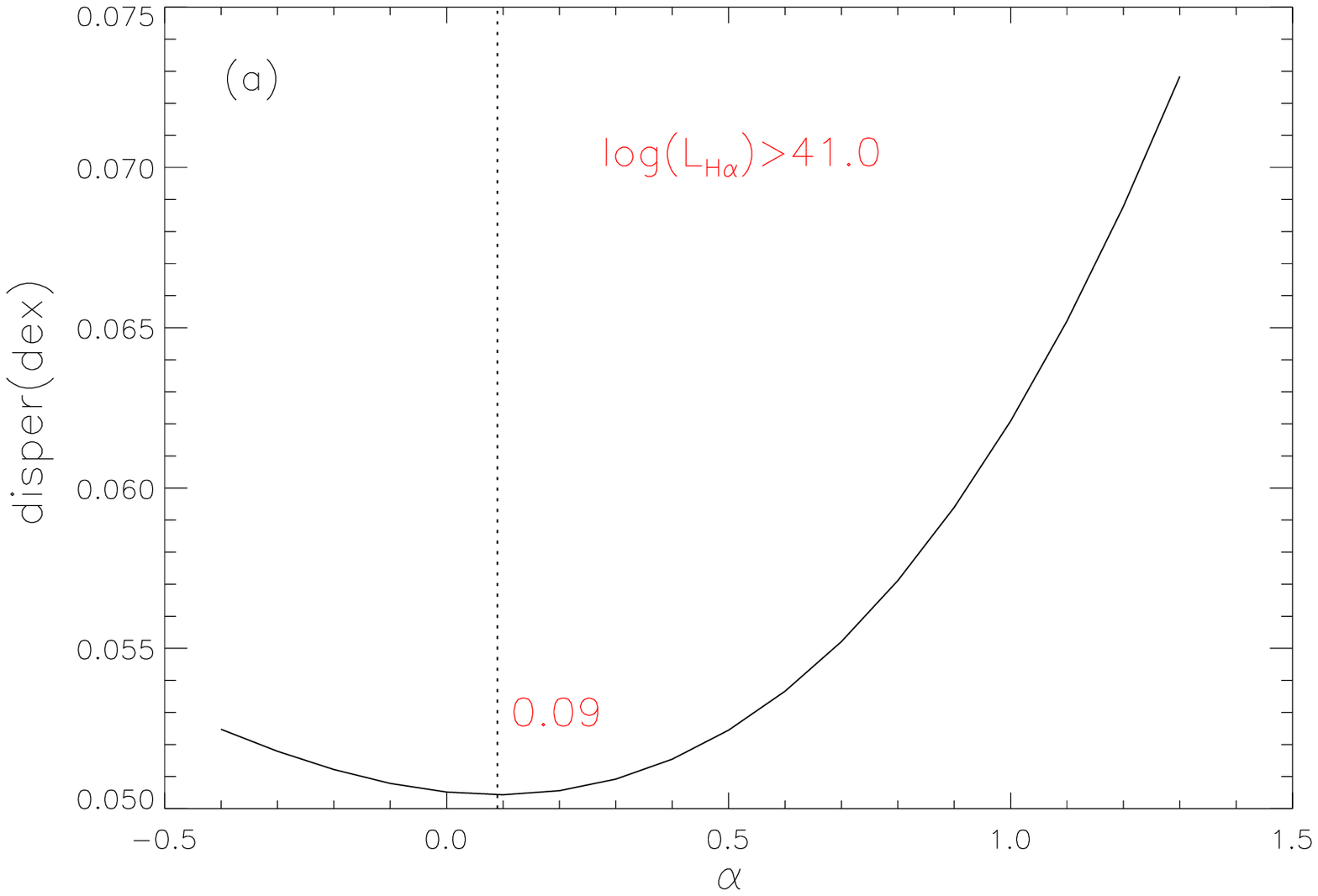}
\includegraphics[width=7cm,height=5.25cm]{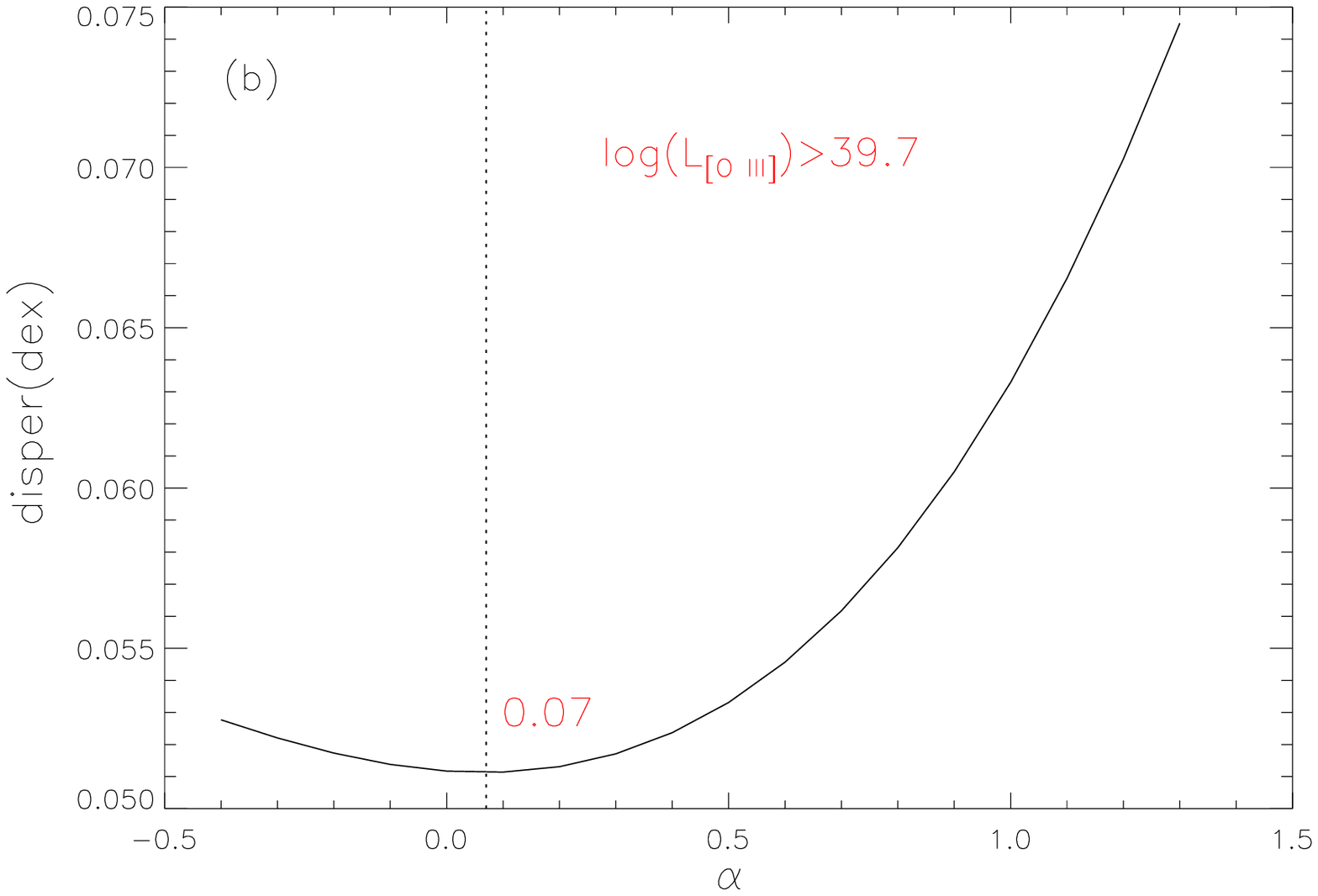}
\caption{Dispersion of metallicities as a function of $\alpha$. The
$\alpha$ values show a weak dependence on SFR. Left-hand panel:
log($ L_{\rm H\alpha})>41.0$ sample. Right-hand
 panel: log($L_{\rm \oiii})>39.7$ sample.}
\end{center}
\end{figure*}

Considering the luminosity thresholds of log$(L_{\rm H
\alpha})>41.0$ and log$(L_{\rm \oiii})>39.7$, we show respectively
the MZ relations with the typical error bars of $0.10$ dex and
$0.11$ dex in Figures 3(a) and 3(b). In Figure 3(a), we sort SFGs of
each subsample into 16 bins with the same galaxy stellar mass ranges
as in Figure 1, including more than 10 galaxies in each bin. All
median values of stellar masses and metallicities of SFGs in each
bin are shown in Figure 3(a). In addition, we employ the
second-order polynomial to fit these data of each subsample, and
present them with different lines. In Figure 3(a), the metallicity
difference of MZ relations between the lowest and highest redshift
ranges at $\rm log (M_{*}/M_{\sun})\sim9.3$ is about $0.15$ dex (the
red and black lines) in Figure 3(a). At
$\rm~log(M_{*}/M_{\sun})<10.0$, Figure 3(a) agrees with the result
of Figure 1, showing metallicity evolution; at
$\rm~log(M_{*}/M_{\sun})>10.0$, the evolution is not observed in
Figure 3(a), and the evolution observed in Figure 1 should have an
observational bias.

Following Figure 3(a), we present them by employing the luminosity
threshold of log$(L_{\rm \oiii})>39.7$ in Figure 3(b). Compared with
Figure 3(a), the SFG sample of Figure 3(b) increases by a factor of
two approximately; therefore, their median metallicities and stellar
masses in each redshift subsample are more smoothly distributed.
Like Figure 3(a), Figure 3(b) also presents the MZ relation
evolution, with the maximum metallicity difference of $\sim 0.15$
dex at $\rm~log(M_{*}/M_{\sun})<10.0$; Figure 3(b) agrees almost
with the result of Figure 3(a) at $\rm~log(M_{*}/M_{\sun})>10.0$. In
addition, we find that the evolution of the MZ relation seems to
disappear at about $\rm log(M_{*}/M_{\sun})>10.0$ after applying the
two luminosity thresholds. The reasons may be (1) oxygen enrichment
decreases slowly with increasing galaxy stellar mass (Figure 3 of Wu
\& Zhang 2013), presenting the galaxy downsizing effect (Pilyugin \&
Thuan 2011); (2) metallicity saturates firstly at more massive
galaxies and then at less massive galaxies (Zahid et al. 2013); our
SFG sample has a lower range of $0.04 < z < 0.12$, with the median
redshift of $0.076$, and therefore the metallicity saturation may
begin at $\rm log(M_{*}/M_{\sun}) \sim 10.0$.

In Table 2, we show the KS-test probabilities of the same
metallicity distributions for Figures 3(a) and 3(b) between
$0.04<z\leqslant0.06$ and $0.10<z<0.12$ subsamples at the mass bins
centered (median values of $\rm log(M_{*}/M_{\sun})$ on $\sim9.3$,
$\sim9.5$, $\sim9.65$, $\sim9.8$, $\sim9.9$, and $\sim10.1$; the
median masses between the $0.10<z<0.12$ and $0.04<z\leqslant0.06$
subsamples differ less than 0.05 in each mass bin. A K-S test shows
that the probabilities, the same parent population between the two
redshift subsamples, are smaller and smaller with decreasing mass
bins centered (see Table 2). In addition, we also obtain the 2-d KS
probabilities that the metallicity and stellar mass (their median
values) between the higher redshift (i.e., $0.06<z\leqslant0.07$ ,
$0.07<z\leqslant0.08$, $0.08<z\leqslant0.10$, and $0.10<z<0.12$) and
$0.04<z\leqslant0.06$ subsample SFGs are drawn from the same
metallicity and stellar mass distributions in Figure 3(a), with
0.83606, 0.29806, 0.05839, and 0.00684, respectively. Figure 3(b)
also presents the similar result of Figure 3(a). These suggest that
Figures 3(a) and 3(b) show surely the metallicity evolution between
$0.10<z<0.12$ and $0.04<z\leqslant0.06$ subsample SFGs, indicating
that the redshift evolution of the MZ relation is only evident at
$\rm log(M_{*}/M_{\sun})<10.0$ in the local data.

\begin{table}
\caption{The KS-test results of the local SFGs.}
\begin{small}
\begin{center}
\setlength{\tabcolsep}{10.5pt}
\renewcommand{\arraystretch}{1.2}
\begin{tabular}{ccccl}
\hline \hline
 $\rm log(M_{*}/M_{\sun})$ &$p_{null_{a}}$ & $p_{null_{b}}$ & \\
(1)& (2) & (3) & \\
\hline

  $\sim9.3$  & $1.19\times10^{-5}$   & $3.8\times10^{-25}$ & \\
  $\sim9.5$  & 0.013                 & $1.2\times10^{-14}$ & \\
  $\sim9.65$ & 0.005                 & $5.4\times10^{-11}$  & \\
  $\sim9.8$  & 0.289                 & $6.0\times10^{-9}$  & \\
  $\sim9.9$  & $5.84\times10^{-4}$   & $3.6\times10^{-5}$  & \\
  $\sim10.1$ & 0.172                 & 0.16                & \\
\hline \hline

\end{tabular}
\parbox{3.0in}
{\baselineskip 11pt \noindent \vglue 0.5cm {\sc Note}: Col.(1):
median of log$(M_{*}/M_{\sun})$. Col.(2)-(3). their probabilities
($p_{null_{a}}$ and $p_{null_{b}}$ for Figures 3(a) and 3(b),
respectively.) for the null hypothesis that the
$0.04<z\leqslant0.06$ and $0.10<z<0.12$ distributions are randomly
drawn from the same metallicity population.}
\end{center}
\end{small}
\end{table}

\subsection{Investigating SFR Dependance}

In Section 3.1, we have presented the evolution of the MZ relation.
Since the evolution remains controversial, we explore whether the
evolution depends on SFRs. For this aim, we use the method of
Mannucci et al. (2010) to investigate SFR dependence.

Corresponding to Figures 3(a) and 3(b), we respectively present the
relations of stellar masses and SFRs (M-SFR) with their typical
error bars of 0.17 dex and 0.19 dex in Figures 3(c) and 3(d). All
median values of stellar masses and SFRs of SFGs in each bin are
shown in Figures 3(c) and 3(d). In addition, we employ the
second-order polynomial to fit these data of each subsample, and
present them with different lines. Figure 3(d) shows significantly
SFR evolution, whereas the evolution is not observed in Figure 3(c).
There may be two reasons: (1) SFRs are perhaps not the main driving
force of metallicity evolution, since metallicity evolves more with
the stellar mass than with the SFR in the SDSS data (Mannucci et al.
2010); (2) when log$(L_{\rm H\alpha})>41.0$ is used as the minimum
luminosity threshold, galaxies with similarly high SFRs at all
redshifts are selected (see Figures 3(c) and 4(a)), and therefore no
SFR evolution is observed; when log$(L_{\rm \oiii})>39.7$ is used as
the minimum luminosity threshold, galaxies at all redshifts are
selected randomly (see Figures. 3(d) and 4(b)), and therefore SFR
evolution is observed.

Next, we will investigate why a similar metallicity offset is shown
in Figures 3(a) and 3(b) despite the fact that the SFR evolution is
different in the log$(L_{\rm H\alpha})>41.0$ and log$(L_{\rm
\oiii})>39.7$ selected samples. Figures 4(a) and 4(b) present the
contours of SFRs used as functions of log($\rm H\alpha$) and
log($\rm \oiii$) luminosities; Figures 4(c) and 4(d) respectively
show the contours of 12+log(O/H) used as functions of log($\rm
H\alpha$) and log($\rm \oiii$) luminosities. Although the
significantly different log(SFR) distributions between the two
minimum thresholds are observed (see Figures 4(a) and 4(b)), SFGs in
each subsample have a similar metallicity distribution, from
12+log(O/H) $\sim 8.5$ to $\sim 9.2$ (see Figures 4(c) and 4(d)),
when the observed minimum line luminosity is either log$(L_{\rm H
\alpha})>41.0$ or log$(L_{\rm \oiii})>39.7$. This explains the
similar metallicity offset in Figures 3(a) and 3(b) under the
different luminosity thresholds. These indicate that metallicity
evolution is shown well under the luminosity thresholds of
log$(L_{\rm H \alpha})>41.0$ or log$(L_{\rm \oiii})>39.7$, and that
SFR evolution still is shown well under the latter luminosity
threshold, but the evolution is not observed under the former
luminosity threshold.

Here, we measure directly whether there is an SFR dependence using
both log$(L_{\rm H \alpha})>41.0$ and log$(L_{\rm \oiii})>39.7$
selected samples. Following the method of Mannucci et al. (2010), we
investigate the $\alpha$ value in the equation ($\mu={\rm
log}M_{*}-\alpha \rm log(SFR)$) which minimizes the metallicity
scatter in bins of mass and SFR for both samples. Figures 5(a) and
5(b) show respectively $\alpha =0.09$ and $\alpha =0.07$ for
log$(L_{\rm H \alpha})>41.0$ and log$(L_{\rm \oiii})>39.7$ samples,
and therefore these exclude fully a strong dependence of MZ relation
on SFR, suggesting a weaker dependence on SFR, and the two $\alpha$
values are close to the $\alpha=0.05$ value of De Los Reyes et al.
(2015). Moreover, the metallicity difference between $\alpha=0.09$
or $0.07$ and $\alpha=0$ is about 0.0001 dex, and therefore this may
be further evidence for the weak SFR dependence.

\subsection{Dependance of metallicity evolution on aperture
 covering fraction}

\begin{figure}
\begin{center}
\includegraphics[width=8cm,height=6cm]{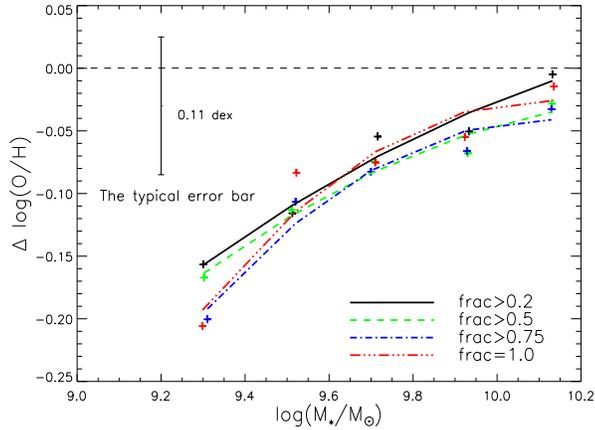}
\caption{The metallicity differences for the different covering
fractions from SFGs at $0.04<z<0.12$, considered the minimum
luminosity limit of log$(L_{\rm \oiii})>39.7$. The metallicity
difference is the same as in Figure 3(b). $0.11$ dex in Figure 7 is
the typical error bar.}
\end{center}
\end{figure}

In Section 3.1, we have shown that the redshift evolution of
metallicities, at least for log$(M_{*}/M_{\sun})<10.0$, remains
after accounting for line luminosity selection effects. In this
section, as a final check, we explore an alternative bias which may
mimic redshift evolution, namely the aperture covering fraction.

In the sample of SDSS DR7 data, we employ the redshift limit of
$0.04<z<0.12$. Following the suggestion of Kewley et al. (2005), we
require the lower redshift limit of 0.04. In addition, the aperture
covering fraction of $> 20\%$ is required. Using the upper limit of
$z<0.12$ (Zahid et al. 2013, 2014), we mainly focus on minimizing
evolution effect to avoid a select effect of metallicity evolution
originating from higher redshift galaxies. The redshift and aperture
covering fraction limits should avoid or reduce the biases from the
aperture effect and selection effect. Therefore we can obtain
reliably the global metallicity estimates for the 3 arcsec SDSS
fiber.

In Figure 1, the metallicity differences between higher redshift and
lower redshift subsamples are $\sim 0.1-0.2$ dex. Although negative
metallicity gradients (i.e. cores with higher metallicity than
outskirts) are presented in most galaxies (McCall et al. 1985;
Vila-Costas \& Edmunds 1992; Henry \& Worthey 1999; Bresolin et al.
2012), we suggest that the metallicity difference in Figure 1 does
not stem from the metallicity gradients, because the difference
between metallicity gradients of large disks and either small disks
or dwarf galaxies is much less than $0.05~ \rm dex ~kpc^{-1}$
(Ellison et al. 2008). Then, Ellison et al. (2008) empirically
checked the metallicity gradients by using cuts for r-band
half-light radii ($r_{h}$) and covering fraction, and they found no
evidence for the difference (the median covering fraction of our
sample is $47\%$, compared with their fraction of $30\%$) derived
from the offset effect.

In addition to the above methods, we investigate the metallicity
evolution between the highest and lowest redshift subsamples by
raising the aperture covering fractions, such as $>50\%$, $>75\%$,
and $=100\%$; the metallicity differences for the different aperture
coverings are shown respectively in Figure 6; their median values
are $\sim 79\%$, $\sim 93\%$, and $\sim 100\%$ in their samples.
Considering the observational bias of the emission line luminosity
limits, Figure 6 shows that the metallicity differences between the
highest and lowest redshfit subsamples for the different aperture
covering fractions, from $> 20\%$ to $> 50\%$, $> 75\%$ and then $=
100\%$ are consistent within the $1\sigma$ error, and their typical
error bar is displayed in Figure 6. Also, the metallicity difference
is $\sim 0.15~$dex at $\rm log (M_{*}/M_{\sun})\sim9.3$. Therefore
we suggest that the metallicity evolution between SFGs with
different redshift ranges is indeed real.

\section{Summary}

Considering the aperture effect and selection effect, we utilized
$0.04<z<0.12$ and aperture covering fractions of $>20\%$ for r band,
and chose galaxies with SNR $>$ 3 for H$\beta~\lambda 4861$,
H$\alpha~\lambda 6563$, $\rm \oii~\lambda \lambda 3727, 3729$, and
$\rm \nii~ \lambda 6584$. Finally, we obtained 53,444 SFGs, selected
from the catalog of MPA-JHU emission-line measurements for the SDSS
DR7. Based on the sample, we investigated the evolution of the MZ
relation. Our main results are the following.

1. In our sample, we showed clearly the redshift evolution of
log($\rm H\alpha$) and log($\rm \oiii$) luminosities. This indicates
that using the SDSS or other data to investigate metallicity
evolution should be corrected by an observed minimum line luminosity
limit.

2. Correcting the observational bias and raising the aperture
covering fractions, we found the metallicity evolution of $\sim
0.15$ dex at $\rm log (M_{*}/M_{\sun})\sim9.3$ in our sample,
indicating that the redshift evolution of the MZ relation is only
evident at $\rm log(M_{*}/M_{\sun})<10.0$ in the local data.

3. After applying the luminosity thresholds of log$(L_{\rm H
\alpha})>41.0$ and log$(L_{\rm \oiii})>39.7$, we found that
metallicity evolution is shown well, and that SFR evolution still is
shown well under the latter luminosity threshold, but the evolution
is not observed under the former one.

4. The evolution of the MZ relation seems to disappear at about $\rm
log(M_{*}/M_{\sun})>10.0$ after applying the luminosity threshold of
log$(L_{\rm H \alpha})>41.0$ or log$(L_{\rm \oiii})>39.7$. This is
due to that (1) oxygen enrichment decreases with increasing galaxy
stellar mass, showing the galaxy downsizing effect; (2) metallicity
saturates firstly at the more massive galaxies and then at the less
massive ones, and therefore the metallicity saturation may begin at
$\rm log(M_{*}/M_{\sun}) \sim 10.0$ in our SFG sample with $0.04 < z
< 0.12$.

5. We explored the $\alpha$ value in the equation ($\mu={\rm
log}M_{*}-\alpha \rm log(SFR)$) which minimizes the metallicity
scatter in bins of mass and SFR following the method of Mannucci et
al. (2010), and found $\alpha =0.09$ and $\alpha =0.07$ for the
log$(L_{\rm H \alpha})>41.0$ and log$(L_{\rm \oiii})>39.7$ samples,
respectively; Moreover, the metallicity difference between the two
$\alpha$ values and $\alpha=0$ is about 0.0001 dex, indicating
further evidence for the weak SFR dependence. These imply that the
MZ relation evolution may have a weaker dependence on SFR in our
sample.

\section*{Acknowledgement}

We appreciate very much the anonymous referee for valuable
suggestions and comments, which allowed us to improve the paper
significantly. SNZ acknowledges partial funding support by the
Strategic Priority Research Program ¡°The Emergence of Cosmological
Structures¡± of the Chinese Academy of Sciences under grant No.
XDB09000000, by 973 Program of China under grant 2014CB845802, by
the Qianren start-up grant 292012312D1117210, and by the National
Natural Science Foundation of China under grant Nos. 11133002 and
11373036. This project was granted financial support from China
Postdoctoral Science Foundation and supported by the Young
Researcher Grant of National Astronomical Observatories, Chinese
Academy of Sciences.

\bsp
\label{lastpage}


\begin{thebibliography}{}
\bibitem[\protect\citeauthoryear{Baldwin}{1981}]{Ba} Baldwin, J. A., Phillips, M. M., \& Terlevich, R. 1981, PASP, 93, 5
\bibitem[\protect\citeauthoryear{Bresolin et al. 2012}{2012}]{B1} Bresolin, F., Kennicutt, R. C., \& Ryan-Weber E. 2012, \apj, 750, 122
\bibitem[\protect\citeauthoryear{Chabrier 2003}{2003}]{Ch0} Chabrier, G. 2003, PASP, 115, 763
\bibitem[\protect\citeauthoryear{Cullen et al. 2014}{2014}]{C1} Cullen, C., Cirasuolo, M., McLure, R. J., Dunlop, J. S., \& Bowler, R. A. A. 2014, \mnras, 440, 2300
\bibitem[\protect\citeauthoryear{De Los Reyes et al. 2015}{2015}]{D1} De Los Reyes, M. A., et al. 2015, AJ, 149, 79
\bibitem[\protect\citeauthoryear{Ellison et al. 2008}{2008}]{e0} Ellison, S. L., Patton, D. R., Simard, L., \& McConnachie, A. W. 2008, \apjl, 672, 107
\bibitem[\protect\citeauthoryear{Erb et al. 2006}{2006}]{e1} Erb, D. K., Shapley, A. E., Pettini, M., et al. 2006, \apj, 644, 813
\bibitem[\protect\citeauthoryear{Foster et al. 2012}{2012}]{F0} Foster, C., et al. 2012, \aap, 547, 79
\bibitem[\protect\citeauthoryear{Henry et al. 2013a}{2013a}]{He0} Henry, A., Martin, C. L., Finlator, K., \& Dressler, A. 2013a, \apj, 769, 148
\bibitem[\protect\citeauthoryear{Henry et al. 2013b}{2013b}]{He1} Henry, A., Scarlata, C., Dom\'{\i}nguez, A., Malkan, M., Martin, C. L., et al. 2013b, \apjl, 776, 27
\bibitem[\protect\citeauthoryear{Henry \& Worthey 1999}{1999}]{He2} Henry R. B. C., \& Worthey G. 1999, PASP, 111, 919
\bibitem[\protect\citeauthoryear{Juneau et al. 2014}{2014}]{J1} Juneau, S., Bournaud, F., Charlot, S., et al. 2014, \apj, 788, 88
\bibitem[\protect\citeauthoryear{Kauffmann et al. 2003}{2003}]{Ke2} Kauffmann, G., Heckman, T. M., Tremonti, C., et al. 2003, \mnras, 346, 1055
\bibitem[\protect\citeauthoryear{Kewley \& Ellison 2008}{2008}]{Ke0} Kewley, L. J., \& Ellison, S. L. 2008, \apj, 681, 1183
\bibitem[\protect\citeauthoryear{Kewley et al. 2006}{2006}]{Ke1} Kewley, L. J., Grovers, B., Kauffmann, G., \& Heckman, T. 2006, \mnras, 372, 961
\bibitem[\protect\citeauthoryear{Kewley et al. 2005}{2005}]{Ke3} Kewley, L. J., Jansen, R. A., \& Geller, M. J. 2005, PASP, 117, 227
\bibitem[\protect\citeauthoryear{Kroupa 2001}{2001}]{Kr0} Kroupa, P. 2001, \mnras, 322, 231
\bibitem[\protect\citeauthoryear{Lara-L\'{o}pez et al. 2010}{2010}]{La0} Lara-L\'{o}pez, M. A., Bongiovanni, A., et al. 2010, \aap, 519, 31
\bibitem[\protect\citeauthoryear{Lara-L\'{o}pez et al. 2013}{2013}]{La1} Lara-L\'{o}pez, M. A., Hopkins, A. M., L\'{o}pez-S\'{a}nchez, A. R., et al. 2013, \mnras, 434, 451
\bibitem[\protect\citeauthoryear{Lequeux et al. 1979}{1979}]{Le1} Lequeux, J., Peimbert,M., Rayo, J. F., Serrano, A., \& Torres-Peimbert, S. 1979, \aap, 80, 155
\bibitem[\protect\citeauthoryear{Lian et al. 2015}{2015}]{Li0} Lian, J. H., Li, J. R., Yan, W., Kong, X. 2015, \mnras, 446, 1449
\bibitem[\protect\citeauthoryear{Maiolino et al. 2008}{2008}]{Ma2} Maiolino, R., Nagao, T., Grazian, A., et al. 2008, \aap, 488, 463
\bibitem[\protect\citeauthoryear{Mannucci et al. 2009}{2009}]{Ma4} Mannucci, F., Cresci, G., Maiolino, R., et al. 2009, \mnras, 398, 1915
\bibitem[\protect\citeauthoryear{Mannucci et al. 2010}{2010}]{Ma5} Mannucci, F., Cresci, G., Maiolino, R., Marconi, A., \& Gnerucci, A. 2010, \mnras, 408, 2115
\bibitem[\protect\citeauthoryear{McCall et al. 1985}{1985}]{Mc0} McCall M. L., Rybski P. M., \& Shields G. A. 1985, \apjs, 57, 1
\bibitem[\protect\citeauthoryear{Noeske et al. 2007}{2007}]{N1} Noeske, K. G., Weiner, B. J., Faber, S. M., et al. 2007, \apjl, 660, L43
\bibitem[\protect\citeauthoryear{Pilyugin et al. 2006}{2006}]{Pi1} Pilyugin, L. S., Thuan, T. X., V\'{\i}lchez, J. M. 2006, \mnras, 367, 1139
\bibitem[\protect\citeauthoryear{Pilyugin et al. 2010}{2010}]{Pi2} Pilyugin, L. S., V\'{\i}lchez, J. M., Cedr\'{e}s, B., \& Thuan, T. X. 2010, \mnras, 403, 896
\bibitem[\protect\citeauthoryear{Pilyugin \& Thuan 2011}{2011}]{Pi3} Pilyugin, L. S., \& Thuan, T. X. 2011, \apj, 726, L23
\bibitem[\protect\citeauthoryear{Salim et al. 2015}{2015}]{Sa0} Salim, S., Lee, J. C., Dav\'{e}, R., Dickinson, M. 2015, \apj, 808, 25
\bibitem[\protect\citeauthoryear{Salim et al. 2007}{2007}]{Sa2} Salim, S., Rich, R. M., Charlot, S., et al., 2007, \apjs, 173, 267
\bibitem[\protect\citeauthoryear{Savaglio et al. 2005}{2005}]{Sa2} Savaglio, S., Glazebrook, K., Borgne, D. L., Juneau, S., Abraham, R. G., et al. 2005, \apj, 635, 260
\bibitem[\protect\citeauthoryear{Steidel et al. 2010}{2010}]{Se0} Steidel, C. C., Erb, D. K., Shapley, A. E., et al. 2010, \apj, 717, 289
\bibitem[\protect\citeauthoryear{Steidel et al. 2014}{2014}]{Se1} Steidel, C. C., Rudie, G. C., Strom, A. L., et al. 2014, \apj, 795, 165
\bibitem[\protect\citeauthoryear{Tremonti et al. 2004}{2004}]{Tr1} Tremonti, C. A., Heckman, T. M., Kauffmann, G., et al. 2004, \apj, 613, 898 (T04)
\bibitem[\protect\citeauthoryear{Vila-Costas \& Edmunds 1992}{1992}]{vi0} Vila-Costas M. B., \& Edmunds M. G. 1992, \mnras, 259, 121
\bibitem[\protect\citeauthoryear{Weiner et al. 2009}{2009}]{We1} Weiner, B. J., Coil, A. L., Prochaska, J. X., et al. 2009, \apj, 692, 187
\bibitem[\protect\citeauthoryear{Whitaker et al. 2012}{2012}]{Wh1} Whitaker, K. E., van Dokkum, P. G., Brammer, G., \& Franx, M. 2012, \apjl, 754, 29
\bibitem[\protect\citeauthoryear{Wu et al. 2013}{2013}]{Wu2} Wu, Y.-Z., \& Zhang, S.-N. 2013, \mnras, 436, 934
\bibitem[\protect\citeauthoryear{Yabe et al. 2012}{2012}]{Y0} Yabe, K., Ohta, K., Iwamuro, F., et al. 2012, PASJ, 64, 60
\bibitem[\protect\citeauthoryear{Yabe et al. 2014}{2014}]{Y1} Yabe, K., Ohta, K., Iwamuro, F., et al. 2014, \mnras, 437, 3647
\bibitem[\protect\citeauthoryear{Yates et al. 2012}{2012}]{Y1} Yates, R. M., Kauffmann, G., Guo, Q. 2012, \mnras, 422, 215
\bibitem[\protect\citeauthoryear{Zahid et al. 2014}{2014}]{Z4} Zahid, H. J., Dima, G. I., Kudritzki, R., et al. 2014, \apj, 791, 130
\bibitem[\protect\citeauthoryear{Zahid et al. 2013}{2013}]{Z3} Zahid, H. J., Geller, M. J., Kewley, L. J., et al. 2013, \apjl, 771, 19
\end{thebibliography}
\end{document}